\documentclass[article, twocolumn, prl, showpacs, amssymb, superscriptaddress]{revtex4}
\usepackage{graphicx}
\usepackage{epsfig}
\usepackage{amsmath}
\usepackage{dcolumn}
\usepackage{amsmath}
\usepackage{latexsym}
\usepackage{bm}
\usepackage{bbm}

\begin{document}

\title{Fractional Coulomb blockade in a coupling controlled metallic quantum dot.}
\author{O. Bitton}
\affiliation{Chemical Research Support department, Weizmann Institute of Science, Rehovot, Israel}
\affiliation{The institute of nanotechnology and advanced materials, The Department of Physics, Bar Ilan University, Ramat Gan 52900, Israel}
\author{A. Frydman}
\affiliation{The institute of nanotechnology and advanced materials, The Department of Physics, Bar Ilan University, Ramat Gan 52900, Israel}
\author{R. Berkovits}
\affiliation{The institute of nanotechnology and advanced materials, The Department of Physics, Bar Ilan University, Ramat Gan 52900, Israel}
\author{D.B. Gutman}
\affiliation{The institute of nanotechnology and advanced materials, The Department of Physics, Bar Ilan University, Ramat Gan 52900, Israel}

\begin{abstract}

We use a novel technique to experimentally explore transport properties through a single metallic nanoparticle with variable coupling to electric leads. For strong dot-lead coupling the conductance is an oscillatory function of the gate voltage with periodicity determined by the charging energy, as expected. For weaker coupling we observe the appearance of additional multi-periodic oscillations of the conductance with the gate voltage. These  harmonics correspond to a change of the charge on the dot by a fraction of an electron. This notion is supported by theoretical calculations based on dissipative action theory. Within this framework  the multiple periodicity of the conductance oscillations arises due to  non-pertubative instanton solutions.

\end{abstract}


\date{\today}

\maketitle

Transport through a quantum dot (QD) has been extensively studied in the context of two dimensional electron gas in semiconducting systems (for reviews see \cite{review1,review2,Aleiner_Review,Nazarov_book}).  With nanotechnology advances and growing interest in electronic transport through nano particles and molecular devices  it becomes important to understand the fundamental properties of metallic based QDs.  
A distinct property of QDs is the presence of  electric conductance oscillations as a function of  gate voltage, $V_g$, that originate from the Coulomb blockade (CB) effect. Due to the strong sensitivity to the applied voltage these oscillations 
can be potentially utilized  as building blocks of  a  single electron transistor. 

By controlling  the coupling between the dot and the leads one  can drive  the system 
from  the "closed"  to the "open" dot regime.
The charge confined in a closed dot is quantized and one observes a well pronounced 
set of conductance peaks, separated by the charging energy $E_c$.
Each  peak is  associated with  the change of total charge on the dot by one electron. 
On the other hand, the charge in an open dot is not quantized;
Coulomb effects are suppressed and one observes a weak modulation of the conductance through the dot, with the period $E_c$\cite{matveev,Nazarov_book}.

In the current work we explore the transition between these two regimes. For this we use a unique experimental method that enables us to tune the coupling  between a metallic nanoparticle 
and the leads while controlling $V_g$.
In the intermediate regime the conductance is characterized by a multiple periodicity  as a function of the gate voltage. The additional harmonics are associated with a change of the charge of the dot by a fraction of an electron. 
  
In QDs based on low-carrier-density 2D electron gases, dot-lead coupling can be controlled by applying back gate voltages. A similar technique for controlling the coupling between leads and a high-density metallic QD doesn't exist and modifying the coupling  presents a major challenge. Several methods have been used for studying QDs based on metallic nanoparticles or other nano-objects. These include discontinuous films \cite{ralph1}, electromigration \cite{park}, electrostatic trapping \cite{trapping,kuemmeth},  break-junctions
\cite{breakjunctions} and angle evaporation \cite{klein}. None of these techniques provide a way to control the coupling. As a result one usually ends up with a weakly coupled dot. 

\begin{figure}[h]
\vspace{0cm}
\includegraphics[width=0.5\textwidth]{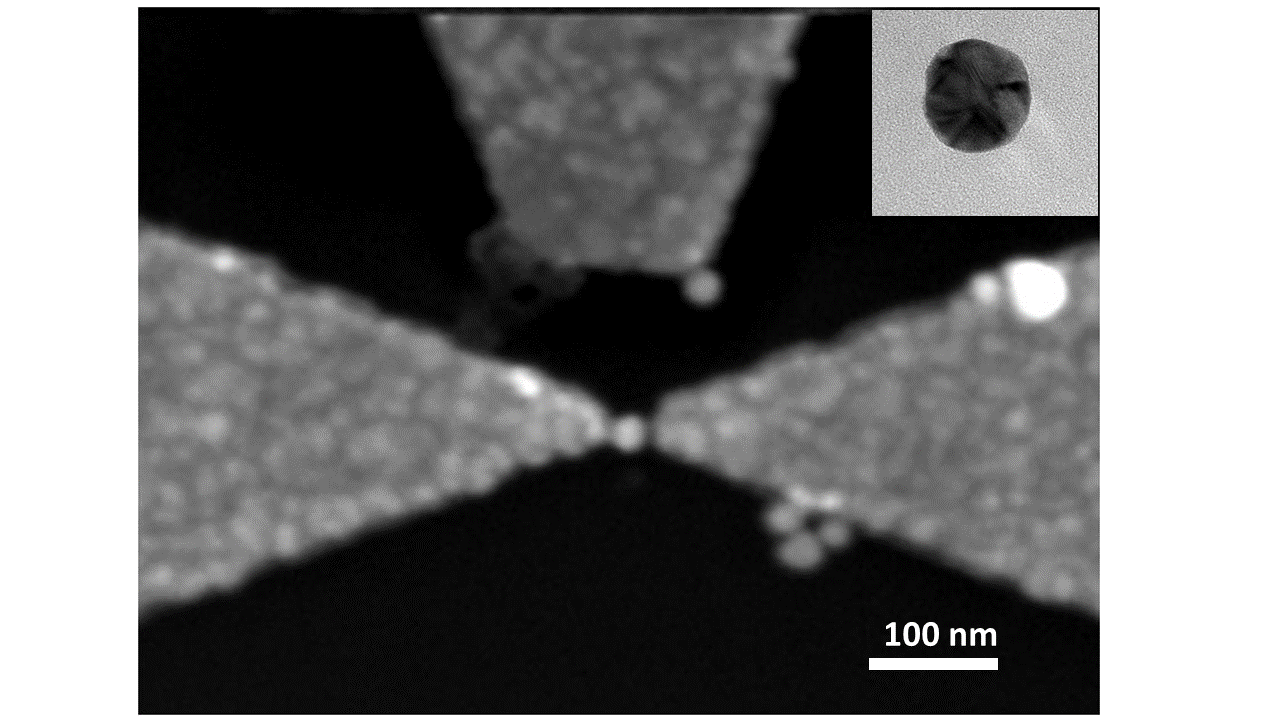}
\caption{ Scanning electron microscopy (SEM) image of a dot-leads system. A 30 nm gold colloid is placed between the source and drain electrodes and a side gate electrode is fabricated 150 nm away. Insert: High resolution TEM image of a gold colloid. \small}
\label{fig1}
\end{figure} 

We have developed a novel technique for fabricating QDs based on metallic nanoparticles, while controllably varying the coupling to leads \cite{liora, liora1}. The quantum dots are chemically formed gold colloids, 30nm in diameter. Coupling to leads is achieved as follows: on a Si-SiO substrate we fabricate two gold electrodes (source and drain) separated by a gap of $10-30 \rm{nm}$ and a perpendicular side gate electrode at a distance of $150 \rm{nm}$. We then electrostatically connect gold colloids to the surface and use Atomic Force Microscope (AFM) nanomanipulation to ``push'' a desired particle to the right position between the source and drain electrodes. At this stage the dot is usually very weakly connected to the leads. We vary the dot-lead coupling using an electrochemical method by which we deposit gold atoms on the gold electrodes to decrease the dot-lead distance \cite{marcus}. During the deposition process we measure the conductance between the source and the drain and stop the process at any desired coupling. We then cool the system to T=4.2K and measure conductance as a function of gate voltage and source drain voltage, $V_{SD}$. Further technical details can be found elsewhere \cite{liora}. An image of a dot-lead system is presented in Fig.\ref{fig1}.

\begin{figure}[h]
\vspace{0cm}
\includegraphics[width=0.55\textwidth]{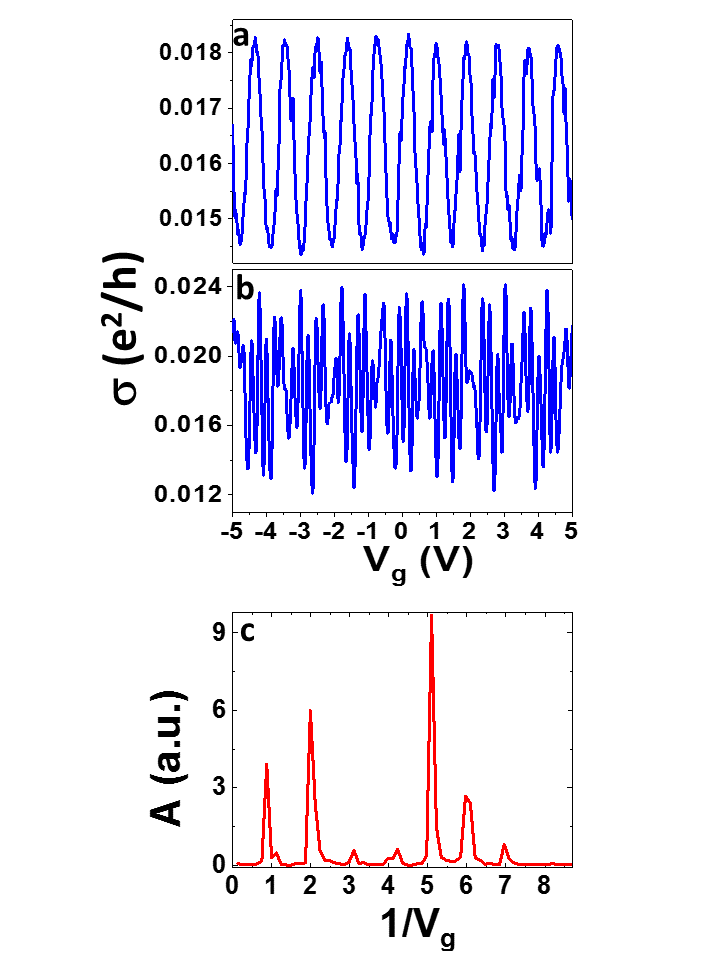}
\caption{ a. $\sigma(Vg)$ for a strongly coupled dot. b. $\sigma(Vg)$ for a weaker coupled dot  c. Fourier transform for the $\sigma(Vg)$ of figure b showing seven harmonics of the coulomb blockade period. \small}
\label{fig2}
\end{figure} 

Conductance versus gate voltage curves $\sigma(V_g)$ for two QDs are shown in Fig.\ref{fig2} panels a and b. In Fig.2a the dot is strongly coupled and the 
conductance oscillates with a single period corresponding to $E_c \sim 0.8V$. In Fig.2b, on the other hand, the dot is weaker coupled and the conductance exhibits a much richer structure.  The Fourier transform depicted in Fig. 2c reveals that the conductance curve is composed of 7 well defined periodicities which are identified as harmonics of the basic CB oscillation which in this case is $\sim 1.8V$. It should be noted that the relative amplitude of the different oscillations is not monotonic with the harmonic order. In this case the second harmonic has a larger amplitude than the first and the fifth harmonic is the most prominent. 

We have measured 17 samples, out of which seven showed several harmonics of $E_c$. It turns out that probability to observe multiple harmonics depends on the dot-lead coupling, characterized by the dimensional conductivity $g_D=(\hbar/e^2)(R_S^{-1}+R_D^{-1})$, where  $R_{S/D}$ is the resistance 
between the dot and source/drain. Unfortunately, it is not possible to measure  $g_D$ directly from the conductivity because the dot is usually highly asymmetrically coupled. While the measured conductivity is governed by the weakly connected lead, $g_D$ is determined by the the well connected lead. Nevertheless it is possible to extract $g_{D}$ from I-V curves, such as that shown in the inset of Fig.\ref{fig2} \cite{liora1}, by fitting them to the results of Ref. \cite{Zaikin}. Doing so we find that the coupling of all our samples is  
in the range  $1<g_{D}<10$. Fig.2 shows the number of conductance oscillation periods observed in the system  as  a function of coupling strength $g_{D}$. It is seen that for $g_D>6$ the conductance shows only a single period. Additional harmonics appear for $g_D<6$ and proliferate  as the system is pushed towards lower values of $g_D$.

\begin{figure}[h]
\vspace{0cm}
\includegraphics[width=0.5\textwidth]{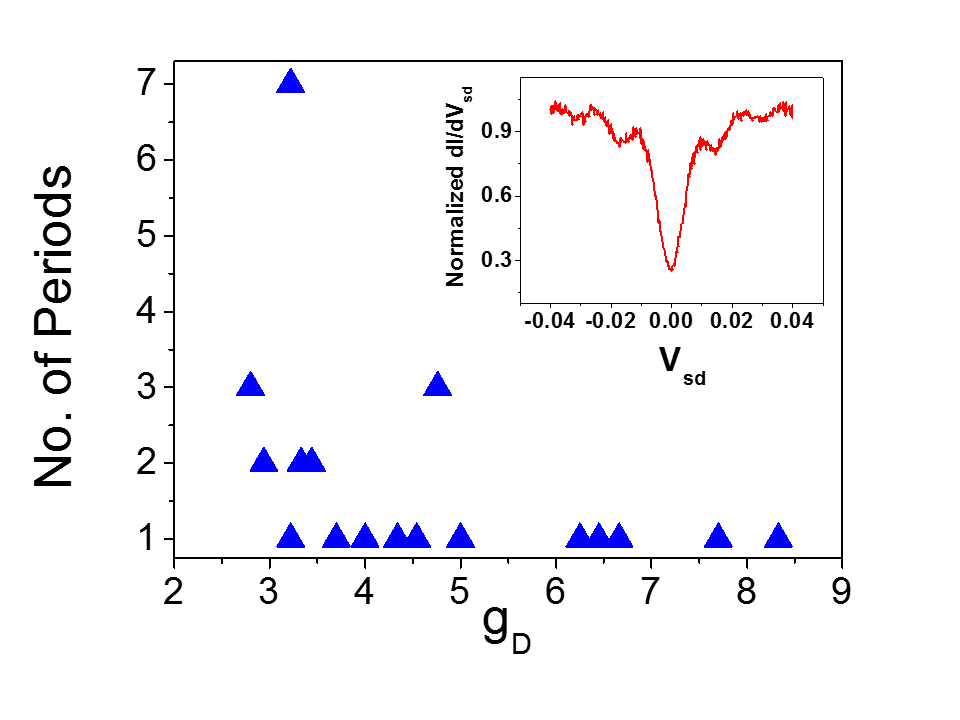}
\caption{ No. of periods as a function of the conductance to the strongly connected lead, $g_{D}$. Insert: Typical $dI/dV_{SD}$ curve for strongly coupled dots.   \small}
\label{fig3}
\end{figure}

\begin{figure}[h]
\vspace{0cm}
\includegraphics[width=0.5\textwidth]{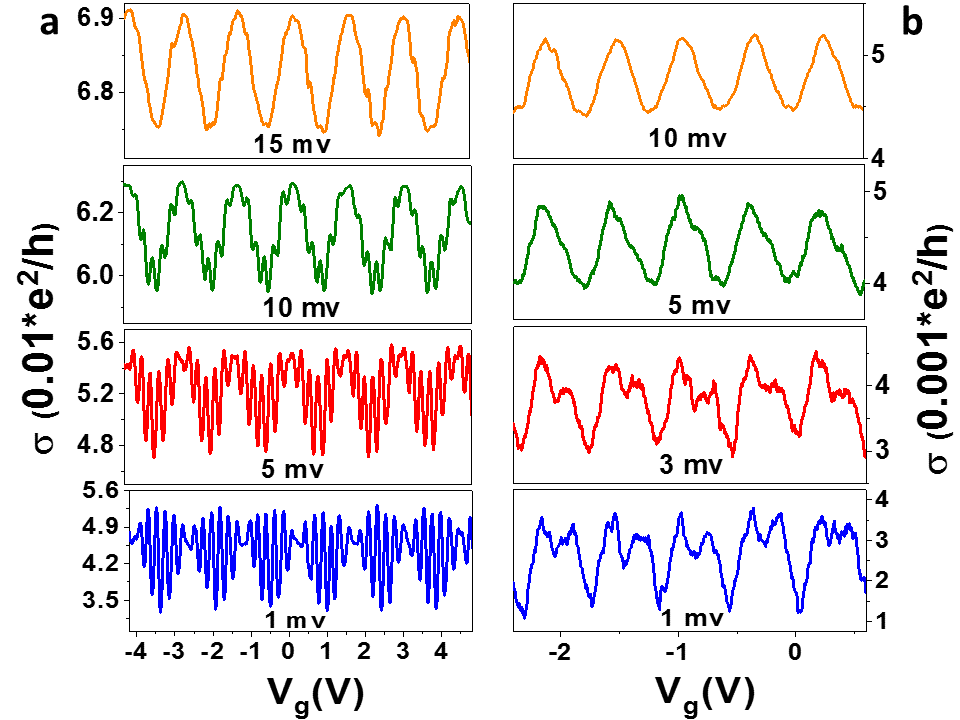}
\caption{ $\sigma(Vg)$ for two different dots at different values of bias voltage: a. The dot is characterized by $g_{D}=4.76$. The ratio between the two periods is 6.  b. The dot is characterized by  $g_{D}=3.44$. The ratio between the two periods is 2.  \small}
\label{fig4}
\end{figure} 

We also note that high and low  harmonics are differently affected by external bias.  
As the bias voltage is increased, the additional harmonics are suppressed. This is demonstrated in Fig.\ref{fig4} which shows  $\sigma(V_g)$ for two different dots at several values of bias voltage. While for small bias voltage an extra harmonic is clearly observed in the conductance curves, at higher $V_{SD}$ this harmonic is unmeasurable. 

To get a better understanding of the behaviour of our system we use the following theoretical model.  We assume that the dynamics inside the dot is chaotic\cite{Alhassid_review}. The coupling of the dot to the leads is achieved  by a large number of 
closed channels, so that the total coupling strength $g_D$ is large ($g_D \gg 1$).
Because  the charge  inside the dot  strongly  fluctuates, 
it is convenient to use  a canonically conjugated variable - phase $\phi$ \cite{Zaikin_review}.
The QD  is thus described  by  a "dissipative action"\cite{Zaikin,Zaikin_review,Burmistrov,Altland,Nazarov}.
At equilibrium, at  temperature $T$,  it is governed by imaginary time action
%\newline
%\begin{widetext}
\begin{eqnarray}&&
\label{action}
S=\frac{gT^2}{4}\int_0^{1/T} d\tau_1d\tau_2\frac{e^{i\phi(\tau_1)-
i\phi(\tau_2)}}{\sinh^2 \pi T(\tau_1-\tau_2)} \nonumber \\&&
+\frac{1}{4E_c}\int_0^{1/T} d\tau\dot{\phi}^2-iq\int_0^{1/T} d\tau\dot{\phi}\, .
\end{eqnarray}
%\end{widetext}
Here $C$ is  the  capacitance of the dot that  is determined 
by capacitive coupling to the source, drain, and the gate,  $C=C_S+C_D+C_g$. 
The average number of electrons on the dot is
\begin{equation}
q=\frac{C_SR_S-C_DR_D}{e(R_S+R_D)}V_{SD}+\frac{C_G}{e}V_g\,.
\end{equation}
%\begin{equation}
%\alpha(\tau)=\frac{T^2}{\sinh^2 (\pi T\tau)}
%\end{equation}
The action (\ref{action}) has non-trivial minima 
\begin{equation}
e^{i\phi(\tau)}=\prod_{n=1}^{|W|}
\bigg(\frac{e^{2\pi i \tau}-z_n}{1-\bar{z}_ne^{2\pi i \tau}}\bigg)^{{\rm sgn }W},
\end{equation}
known as Korshunov instantons\cite{Korshunov}; $z_n$ and $\bar{z}_n$ are global variables that determine 
the position and  the size of the instanton; $W$ is a winding number
that counts   the number of times the phase $\phi$ circles around the origin.
Above a certain temperature (of the order $T_*\simeq E_c\exp[-g_D]$)  the  instantons are rare,
and they independently contribute to the conductance\cite{Altland}
\begin{equation}
\label{conductance}
G(T)\simeq \frac{R_D R_S}{(R_S+R_D)^2}\bigg[
\frac{e^2}{\hbar}\tilde{g}+\sum_{W=1}^\infty a_W
e^{-F_W(T)}\cos(2\pi q W)
\bigg].
\end{equation}
The first term corresponds to the topologicaly trivial contribution with the winding number $W=0$. It  accounts  for renormalization of the Drude conductance due to  the zero bias anomaly  around trivial minima ($\phi$ being a constant).  It results in  replacement of the coupling strength  
by the renormalized  value
$g_D\rightarrow \tilde{g_D}=g_D+\ln(1+\omega^2 t_c^2)$, 
controlled by the infrared energy scale $\omega ={\rm max} (T, eV)$ and
$RC$ time $t_c=\frac{R_sR_D}{R_S+R_D}C$.
The detailed study of zero bias anomaly  in this geometry was performed in \cite{liora1}.
The topologically non-trivial solutions give rise to various harmonics 
in oscillations of conductance  with the gate voltage. The instantons with the winding number $W$ correspond to a harmonic $W$ in the conductance oscillation.   
In the open dot limit, only the first harmonic survives, leading to weak single period oscillation \cite{Goppert}. As the coupling of the dot decreases,  a finite number of harmonics is observed.
For  $g$ of the order  unity an infinite  number of  harmonics appear with a parametrically equal  magnitude, 
and the instanton expansion breaks down.  
This result merges with the one  known  for the strong Coulomb blockade regime\cite{Aleiner_Review}, i.e. 
for $g \ll 1$. 
In this case the conductance is 
\begin{equation}  
\label{strong_blocade}
G \simeq (R_S+R_D)^{-1}\frac{\Delta E/2T}{\sinh \Delta E/2T}
\end{equation}
where $\Delta E=E_c (q-[q])$ is the deviation from the degeneracy point.
At $T \ll E_c$ it corresponds to the sequence of well resolved CB peaks, that in a Fourier space
gives rise to a large number of harmonics  with approximately  same magnitude.
 
For the open dot limit  the amplitude of harmonic number $W$ is controlled by
\begin{equation}
\label{amplitude}
F_W(T)\simeq \frac{\tilde{g}W}{2}+\frac{\pi^2 T}{E_c}W^2
\end{equation} 
$a_W \simeq g^{W+1} \psi^{(m)}(1)/2\pi$, where $\psi(m)$ is the $m^{th}$ order derivative of the digamma function.
The calculation performed above is valid at equilibrium. 
Though  Korshunov instantons  can be computed  within the real time Keldysh formalism\cite{Titov},
the non-equilibrium generalization remains to be done.
On the phenomenological level, the thermal smearing  out of equilibrium 
is accounted by\cite{Zaikin}  $F(T) \rightarrow F(T) +(2\pi/e^2) W^2\sum_{r=S,D}y_r(x)/R_r$, where
$x_r=R_SR_DR_reVC/(R_S+R_D)^2$, and $y(x)=x\arctan(x)-1/2 \ln(1+x^2)$.

\begin{figure}[h]
\vspace{0cm}
\includegraphics[width=0.6\textwidth]{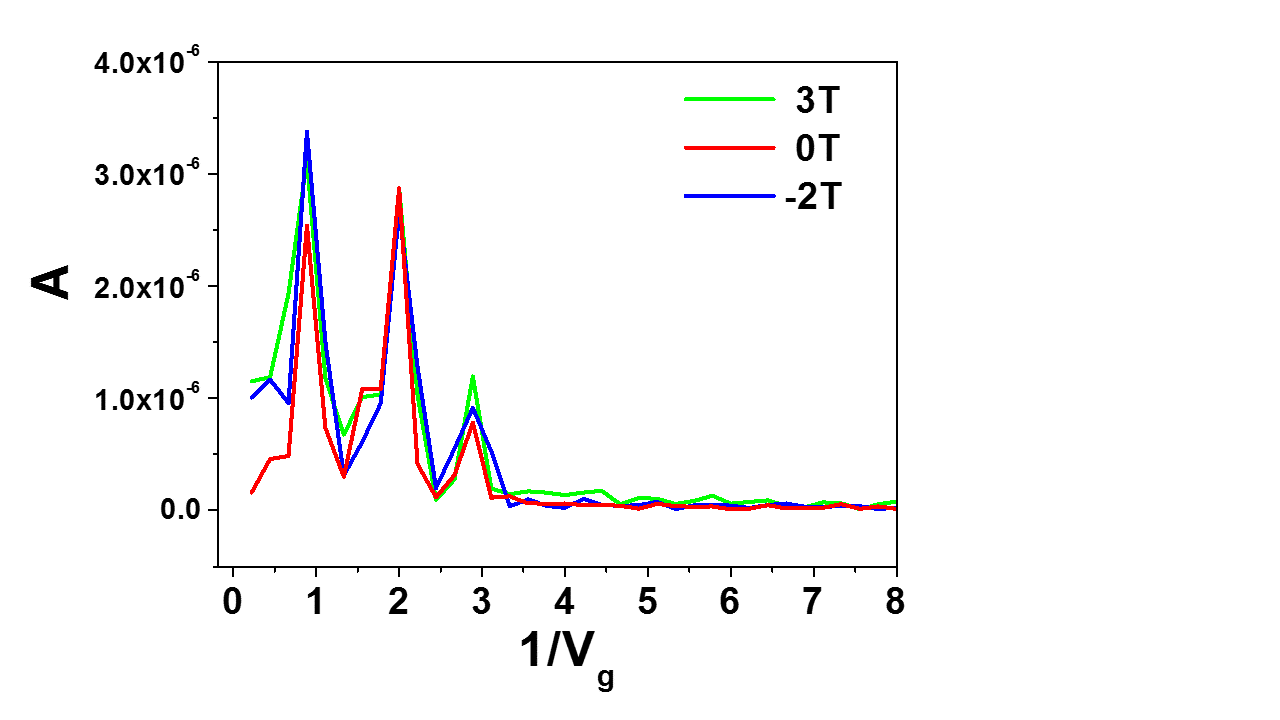}
\caption{ Fourier transform of $\sigma(Vg)$ for a QD at different values of magnetic field with a range corresponding to one flux quantum penetrating the dot (5T).  \small}
\label{fig5}
\end{figure}

Considered as a function of $q$, the  contribution of the instanton with winding number 
$W=1$ is periodic with the period unity,  that corresponds to the charge quantization.
Indeed, changing the average number of electrons in the dot by one, a single periodicity
in the thermodynamic potentials and transport coefficients is expected. 
Instantons with higher winding number give rise to fractional  periodicity that corresponds to an average charge change of $1/W$ of the electron charge.
In an open dot any fraction of the electron can be distributed between inside and outside of the dot, thus periodic dependences with respect to a fraction charge change are possible. 
We believe this picture, i.e. the occurrence of additional harmonics in the conductance is a universal property of open dots, however the weights of the harmonics are model specific. 
For a not-fully-chaotic  dot  the statistics of wave function is not universal, and there is a finite probability to find a strongly coupled states   with a fraction  of an electron inside the dot. 
% can have larger weight than that extracted above for the chaotic case.
Because such  states  give  rise to the pronounced oscillation with a  corresponding harmonic,
one expects  large sample to sample fluctuations  of the harmonic strength.
This may be the reason why  for part  of our samples the strengths  of 
certain high harmonics  is higher than of the lower ones (see Fig.\ref{fig2}). 
This view  is consistent with the observed magnetic field dependence.
Application of  a  magnetic field of the order of a flux quantum through the dot  changes the single particle wave functions, thus affecting the relative magnitudes of different harmonics. Indeed, the Fourier transform shown in Fig.\ref{fig5} for one of our dots reveals that the relative strengths of  various harmonics oscillate with  the magnetic field.  One notes that the  dominant harmonic changes from the first (for $B=-2$T) to the  second  (at  $B=0$T)  and back to the first at (at $B=3$T). The  periodicity range of $5$T corresponds to a single  flux quantum  penetrating the dot.

Within the dissipative action theory the influence of the temperature/bias suppresses the higher harmonics
stronger than the lower ones. It results from the two effects  acting together:
(a) the thermal broadening of the instanton contribution,  Eq.(\ref{amplitude}),  is multiplied by the winding number square ($W^2T/E_c$); 
(b) the  zero bias anomaly  leading  to the logarithmic renormalization of  $\tilde{g_D}$ is weakened  at high temperatures\cite{Grabert,Zaikin_Panyakov}
(compared with  the scale $t_c$), and the terms $\tilde{g_D}W$ increases with temperature.
 Therefore increasing temperature (or voltage) suppresses the harmonics 
with higher winding numbers stronger than those with the lower ones. 
This behavior agrees with  our measurements.  

To conclude,  we have studied the transport through metallic quantum dots while  controllably varying its coupling  to the leads. 
We find  that on the route towards fully opening,  the system passes through  intermediate coupling regime,
in which the conductance shows multi periodic oscillations with the gate voltage. 
We attribute these oscillations to the charging  of the  dot  by a fraction of an electron. 
This notion is supported by calculations based on the dissipative action theory. 

We are grateful for useful discussions with I.S. Burmistrov, Y. Gefen and M. Titov.
This research was supported by the Israeli Science Foundation (grant number 699/13 and 584/14), 
and German-Israeli Foundation (project 1167- 165.14/2011).

\end{document}